\documentstyle[11pt,newpasp,twoside]{article}
\def\edcomment#1{\iffalse\marginpar{\raggedright\sl#1\/}\else\relax\fi}
\reversemarginpar

\begin{document}
\title{XMM-Newton study of the ULIRG NGC~6240}
\author{Ralf Keil, Thomas Boller}
\affil{Max-Planck-Institut f\"ur extraterrestrische Physik,
Postfach 1312, 85741 Garching, Germany}
\author{Ryuichi Fujimoto}
\affil{Institute of Space and Astronautical Science, 3-1-1 Yoshinodai, Sagamihara, Kanagawa 229-8510, Japan}

\begin{abstract}
A recently performed {\em XMM-Newton} observation of the ULIRG NGC~6240 clearly indicates the presence of an AGN contribution to its X-ray spectrum. In the 5.0 - 7.0 keV energy range there is a clear signature of the fluorescent Fe K $\rm \alpha$ lines at 6.4, 6.7 and 6.9 keV, respectively.
 The line strength of the 6.4 keV line cannot be produced by a thermal component. The 0.3 - 10.0 keV spectral energy distribution is characterized by the following components:
(I) two hot thermal components (the starburst),
(II) one direct component (heavily absorbed; AGN is hidden),
(III) one reflection component (the AGN),
(IV) three narrow Fe lines.
The model parameters for the broad-band spectral energy distribution are
consistent with the results of previously works.
\end{abstract}

\vspace*{-1cm}
\section{Introduction}

Many if not all high-luminous infrared galaxies (ULIRGs, $\rm L_{FIR} \ga
10^{12}\, L_{\sun} $) possess regions hidden by huge amounts of dust. This makes it
difficult to ascertain whether this enormous energy output is due to a
starburst activity or an accretion process onto a supermassive black hole.
One of the best known objects to study this relationship is the nearby ULIRG
NGC~6240 (assuming $\rm H_0 \le 65\, km \, s^{-1} \, Mpc^{-1}$). Infrared observations
favour an energy source dominated by starburst processes, whereas observations
in the X-ray range point to an AGN as the central engine ($\rm L_{X} \sim
10^{11}\, L_{\scriptsize\sun}$).

\vspace*{-0.5cm}
\section{Spectral analysis}

We have analyzed the data of NGC~6240 taken from an 24 ksec observation with
{\em XMM-Newton} using the EPIC-PN and EPIC-MOS instruments. In order to
investigate the Fe line complex around 6.4 keV and the 0.3 - 10.0 keV spectrum
as a whole the high sensitivity and therefore the good photon statistics -
especially in the 6.4 keV range - in combination with a higher energy
resolution enables us to examine this feature in unprecedented detail.

\subsection{The Fe K $\rm \alpha$ line complex}
Table 1 summarizes some basic parameters (powerlaw - $\rm \Gamma$, line
energies) of different models (first column) after fitting to the data. The
first of the leading three models includes line profiles with no line width
($\rm \sigma = 0$), whereas eachone of the last two models uses a second
powerlaw, but with a different number of line profiles. Each
model contains a 6.4\,keV line as an indication of an AGN contribution. A prove
of an Compton-thick AGN has been reported by Vignati et al. (1999) using
BeppoSax and by Ikebe et al. (2000) using RXTE. However,
the last model seems to have the best statistical acceptance (see Fig. 1, left).

\begin{table}
\vspace{-0.7cm}
\begin{center}
\begin{scriptsize}
\caption{Spectral fitting results to the Fe line complex}
\begin{tabular}{l|c|ccc|c} \tableline\tableline

Emission lines & powerlaw &\multicolumn{3}{c}{gaussian lines [in keV]} & $\rm \chi^2/$ \\
& $\rm \Gamma$ & Energy-line 1 & Energy-line 2 & Energy-line 3 & d.o.f.\\ \tableline
lines : $\rm \sigma=0$ & -0.18 & $\rm 6.40\,(0.01)$ & $\rm 6.67\,(0.02)$ & $\rm 6.98\,(0.04)$ & 38.5/53\\
lines : $\rm \sigma \ne 0$ & -0.16 & $\rm 6.40\,(0.02)$ & $\rm 6.67\,(0.03)$ & $\rm 6.97\,(0.04)$ & 38.4/51\\
lines : 2nd broad & -0.27 & $\rm 6.40\,(0.03)$ & $\rm 6.60\,(0.11)$ & $\rm 6.66\,(0.06)$ & 43.1/53\\
\tableline\tableline

Emission lines + & powerlaw &\multicolumn{3}{c}{gaussian lines [in keV]} & $\rm \chi^2/$ \\
absorp. edge: & $\rm \Gamma$ & Energy-line 1 & Energy-line 2 & Energy-line 3 & d.o.f.\\ \tableline
po + 2 lines & 0.47 & $\rm 6.39\,(0.02)$ & $\rm 6.65\,(0.04)$ & - & 39.7/54\\
po + 3 lines & 0.47 & $\rm 6.40\,(0.02)$ & $\rm 6.65\,(0.04)$ & $\rm 7.05\,(0.02)$ & 39.1/54\\

\tableline\tableline
\end{tabular}
\end{scriptsize}
\end{center}
\vspace{-0.6cm}
\end{table}

\subsection{Models to the overall X-ray spectrum}
The analysis of the spectral data ($\rm 0.3 - 10.0\,keV$) indicates at least
two models providing an statistically acceptable
fit: Each of them contains two thin thermal plasmas ($\rm kT \approx 0.6\,
keV$ and $\rm kT \approx 1.1\, keV$), a direct component (absorbed powerlaw
with $\rm \Gamma = 1.8$ and $\rm N_H = 2.18 \cdot 10^{24}\,cm^{-2} $, both
fixed) as well as a reflection component (absorbed powerlaw, either reflected
from neutral matter or not). Finally, three gaussian lines have been added to the models (neutral + ionized K $\rm \alpha$ and K $\rm \beta$).
The right plot of Fig. 1 shows the components of the second model (incl. reflection) and
their deviations from the data points. 

\begin{figure}[h]
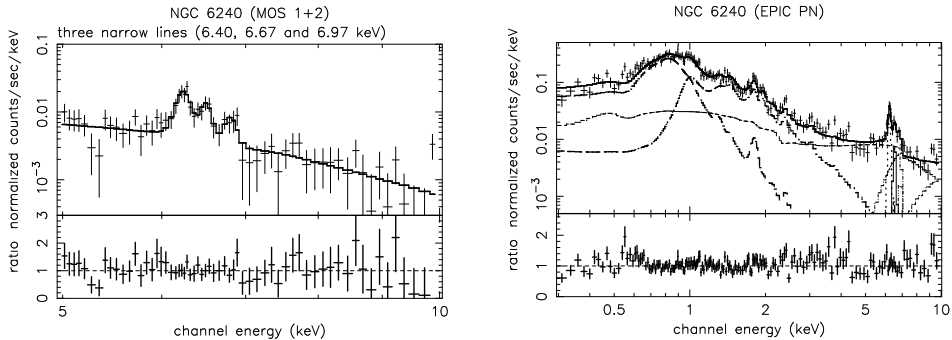

\plotfiddle{Keil_Ralf_fig1.ps}{-1cm}{270}{24}{24}{-180}{25}
\plotfiddle{Keil_Ralf_fig2.ps}{-1cm}{270}{24}{24}{10}{50}
\vspace{2.9cm}
\caption{The Fe K line complex and the overall spectrum}
\end{figure}

The model parameters for the broad-band
spectral energy distribution is consistent with the results previously reported
by Iwasawa \& Comastri (1998), except for the Fe line complex.


\begin{references}
\reference Ikebe, Y. et al. 2000, \mnras, 316, 433
\reference Iwasawa, K. \& Comastri A. 1998, \mnras, 297, 1219
\reference Vignati, P. et al. 1999, \aap, 349, L~57
\end{references}
\end{document}